\documentclass[journal=nalefd,manuscript=letter]{achemso}

\usepackage[T1]{fontenc}
\usepackage{geometry}
\usepackage{setspace}
\usepackage{achemso}
\usepackage{hyperref}
\usepackage{graphicx}
\usepackage{float}
\newfloat{scheme}{htbp}{los}
\floatname{scheme}{Scheme}
\floatname{chart}{Chart}
\newfloat{graph}{htbp}{loh}
\usepackage{chemformula} % Formulas using \ch{}
\usepackage[version = 4]{mhchem} % Formulas using \ce{}
\newcommand{\pin}{\textit{p-i-n} }
\newcommand{\Iwl}{I_{\mathrm{WL}}}
\newcommand{\Iwgm}{I_{\mathrm{WGM}}}
\newcommand{\Itot}{I_{\mathrm{tot}}}
\newcommand{\etawgm}{\eta_{\mathrm{WGM}}}
\newcommand{\upl}{µPL }

\author{Léo~J.~Roche}
\affiliation{Institut für Physik und Astronomie, Technische Universität Berlin, 10623 Berlin, Germany}

\author{Peter~Gschwandtner}
\affiliation{Lehrstuhl für Technische Physik, Physikalisches Institut, Julius-Maximilians-Universität Würzburg, 97074 Würzburg, Germany}

\author{Chirag~C.~Palekar}
\affiliation{Institut für Physik und Astronomie, Technische Universität Berlin, 10623 Berlin, Germany}

\author{Setthanat~Wijitpatima}
\affiliation{Institut für Physik und Astronomie, Technische Universität Berlin, 10623 Berlin, Germany}

\author{Yuhui~Yang}
\affiliation{Institut für Physik und Astronomie, Technische Universität Berlin, 10623 Berlin, Germany}

\author{Monika~Emmerling}
\affiliation{Lehrstuhl für Technische Physik, Physikalisches Institut, Julius-Maximilians-Universität Würzburg, 97074 Würzburg, Germany}

\author{Sven~H\"ofling}
\affiliation{Lehrstuhl für Technische Physik, Physikalisches Institut, Julius-Maximilians-Universität Würzburg, 97074 Würzburg, Germany}

\author{Tobias~Huber-Loyola}
\affiliation{Lehrstuhl für Technische Physik, Physikalisches Institut, Julius-Maximilians-Universität Würzburg, 97074 Würzburg, Germany}

\author{Stephan~Reitzenstein}
\affiliation{Institut für Physik und Astronomie, Technische Universität Berlin, 10623 Berlin, Germany}
\email{stephan.reitzenstein@physik.tu-berlin.de}

\title{Integrated Whispering-Gallery Microlaser–Waveguide Platform for On-Chip Electrical Excitation of InGaAs Quantum Dots}

\begin{document}

\maketitle

\begin{abstract}
We report the fabrication and characterization of an integrated quantum photonic device consisting of an electrically driven whispering-gallery-mode micropillar laser evanescently coupled to a ridge waveguide, both incorporating InGaAs quantum dots (QDs). The lasing characteristics of microlasers are systematically investigated as a function of the pillar–waveguide gap distance. Coherent emission from the whispering-gallery-mode microlaser coupled into the waveguide enables on-chip optical excitation of QDs embedded in an electrically contacted  micropillar at the end of the waveguide.  Under continuous-wave on-chip excitation, we observe single-photon emission with $g^{(2)}(0) = (3.49 \pm 0.01) \%$ for a QD integrated in the outcoupling micropillar which can be spectrally tuned by the quantum confined Stark effect. These results constitute an important step toward low-footprint, deterministic, and scalable single-photon sources for QD-based integrated quantum photonic circuits.
\end{abstract}

%\section*{TOC Graphic}
%\rule{0.05in}{1.75in}%
%\begin{minipage}[b][1.75in]{3.25in}
%  \sffamily
%  \frenchspacing
%	\includegraphics[width=1\textwidth]%{figures/toc/toc.eps}
%\end{minipage}%
%\rule{0.05in}{1.75in}

\section*{Keywords}

on-chip microlaser, single-photon source, quantum dot, whispering gallery mode, integrated quantum photonic circuit

\section{Introduction}

Quantum computing and quantum communication promise a profound paradigm shift in information technology, although their practical realization remains challenged by significant technical and theoretical obstacles. Among the various hardware platforms currently under investigation, solid-state quantum emitters are particularly attractive candidates for photonic quantum computation and are indispensable for encoding so-called flying qubits in quantum communication protocols \cite{Aharonovich2016}. For these applications, non-classical sources must exhibit key characteristics such as on-demand single-photon generation, high brightness and near-unity indistinguishability \cite{Heindel2023}.

While non-deterministic sources based on spontaneous parametric down-conversion (SPDC) can be combined with multiplexing schemes to become quasi-deterministic, they remain less efficient in practice than deterministic single-photon sources \cite{Kaneda2019}. In this context, solid-state quantum dots (QDs) emerge as highly promising candidates \cite{Heindel2023}. These so-called artificial atoms can act as compact and efficient on-demand single-photon emitters. Moreover, their integration in III–V semiconductor materials enables the fabrication of photonic cavities that enhance the spontaneous emission rate via the Purcell effect \cite{Gayral2000, Gayral2008, Hepp2020, Heindel2023}.

With regard to the excitation of single-photon emitters, electrical excitation of QDs via current injection in a \textit{p–i–n} diode structure is intrinsically non-resonant and typically results in limited photon indistinguishability due to charge noise, timing jitter, and photon–phonon interactions \cite{Schlehahn2016}. In contrast, strictly resonant optical excitation of QDs, combined with a stable charge environment, enables the generation of bright single photons with near-unity indistinguishability \cite{Somaschi2016, Reitzenstein2025, Mudi2025}. This excitation scheme is widely regarded as the most effective one for high coherent emission, as it minimizes interactions between the emitted photons and their surrounding environment.

In an ideal integrated quantum photonic circuit (IQPC), each single-photon emitter should be excited independently by a laser. A promising and scalable alternative to bulky external lasers is the integration of on-chip microlasers \cite{Rodt2021}. In particular, whispering-gallery-mode (WGM) micropillar lasers offer several key advantages in this regard. 
They feature lateral emission, low threshold currents and a small footprint, with diameters in the range of a few µm \cite{Albert2010}. Furthermore, due to their in-plane emission they enable the on-chip excitation of photon emitters \cite{Munnelly2017}. Coherent emission from such microlasers has also been shown to couple evanescently into adjacent ridge waveguides \cite{Kryzhanovskaya2023}. 
 
In this work, we present the fabrication and characterization of an IQPC concept in which electrically driven WGM microlasers are integrated with ridge waveguides, enabling evanescent coupling of emission into the waveguides as illustrated in Fig. \ref{fig:layout_schematics_and_sem}(b). After propagation along the waveguide, the laser light excites a Stark-tunable QD integrated into an output micropillar, which generates single photons with high purity. Our optical and quantum-optical investigations include systematic studies of the microlaser performance, the efficiency of light coupling into the waveguide, and the quantum light emission of the QD micropillar integrated at the end of the waveguide.

\section{Device description and fabrication}
\label{sec:device_description_and_fabrication}

To implement our IQPC device concept we use a \textit{p–i–n} diode design along the growth direction to electrically drive the microlaser and to enable spectral fine-tuning of QDs in the waveguide-integrated micropillar (WIM) via the quantum-confined Stark effect (QCSE). For vertical light confinement, the InGaAs QDs are integrated within the intrinsic GaAs cavity, AlGaAs/GaAs distributed Bragg reflector (DBR) pairs are positioned above and below the active region.

\begin{figure}[H]
    \centering
    \includegraphics[width=0.99\textwidth]{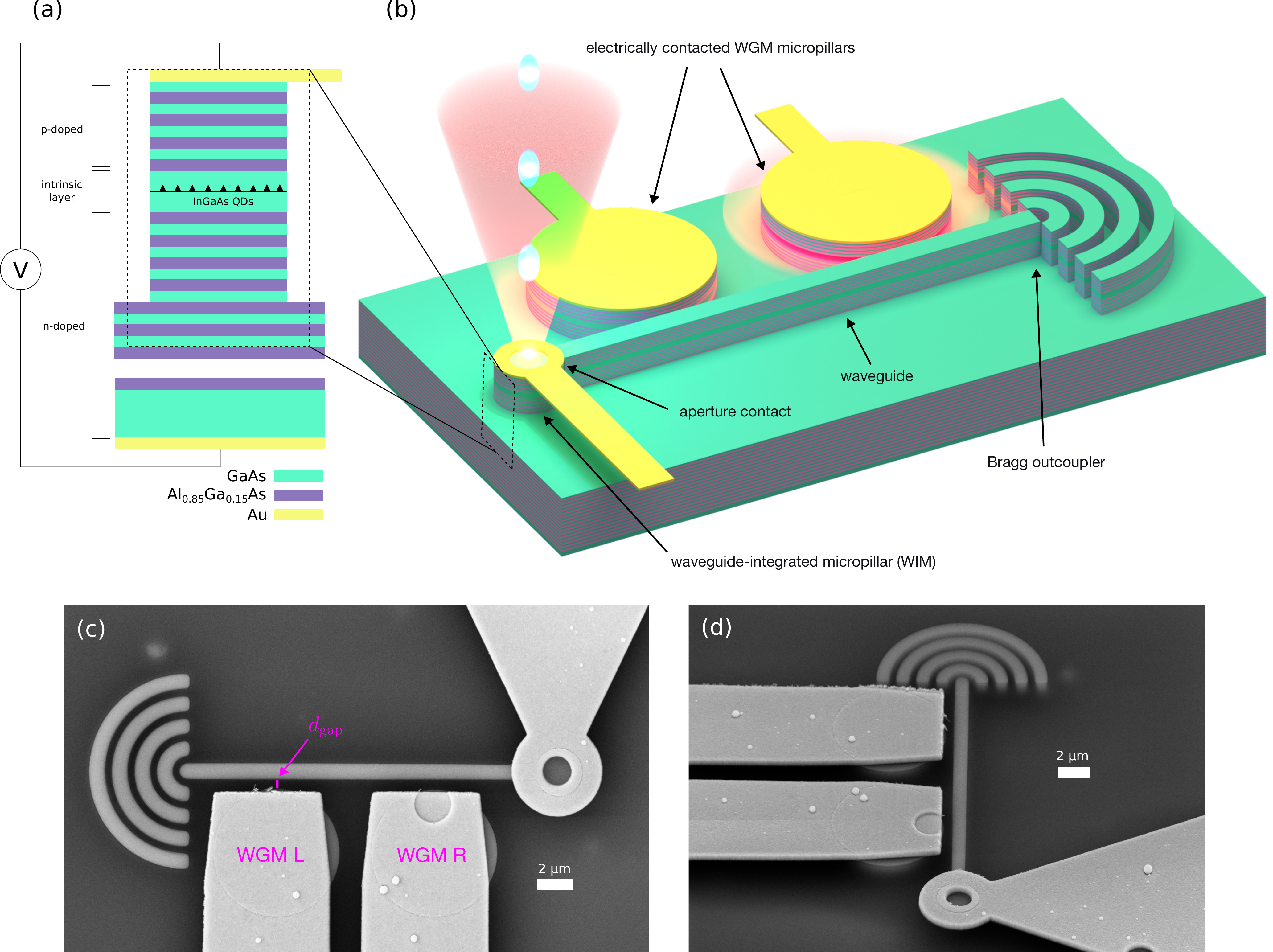}
    \caption{Device architecture and implementation of the monolithic WGM–waveguide–micropillar platform, showing epitaxial \textit{p–i–n} cavity design, on-chip evanescent coupling scheme for coherent excitation and single-photon emission, and SEM images of the fabricated structure. (a) Epitaxial layer structure of the device. 
    (b) Schematic of the monolithic device comprising two electrically contacted WGM micropillars integrated next to a ridge waveguide. One end of the waveguide features a Bragg outcoupler, while the opposite end incorporates an integrated micropillar with an aperture contact for vertical collection of the emitted signal and spectral tuning of the QDs via the QCSE. The red halo indicates coherent emission from a WGM microlaser evanescently coupled into the waveguide, while blue arrows represent single photons emitted by QDs excited on-chip. (c, d) Top-view and high-angle SEM images of the fabricated device. The two WGM micropillars have diameters of 7.2 µm, and $d_{\mathrm{gap}}$ is 500 nm.}
    \label{fig:layout_schematics_and_sem}
\end{figure}

The QD heterostructure was grown by molecular beam epitaxy (MBE) using the layer design depicted in Fig. \ref{fig:layout_schematics_and_sem}(a). On an n-doped GaAs substrate, 20 pairs of $\mathrm{Al}_{0.85}\mathrm{Ga}_{0.15}$As/GaAs distributed Bragg reflector (DBR) layers were deposited, with thicknesses of 79.8 nm and 68.2 nm respectively. The bottom DBRs are Si-doped: the first 15 pairs with a concentration of $3\times10^{18} \mathrm{cm}^{-3}$, and the remaining 5 pairs with $1.5\times10^{18} \mathrm{cm}^{-3}$. Above the bottom DBRs, the intrinsic region was grown, consisting of a single layer of $\mathrm{Al}_{0.85}\mathrm{Ga}_{0.15}$As (79.8 nm) and a GaAs $\lambda$-cavity (272.6 nm), containing a single layer of InGaAs QDs with an estimated density of $10^{19}$ cm$^{-2}$ at its center. On top of the cavity, 4 p-doped DBR pairs with the same layer thicknesses as the bottom DBRs are deposited and C-doped with a concentration of $1\times10^{19} \mathrm{cm}^{-3}$. 

Each fabricated device consists of two electrically contacted 7.2 µm diameter WGM micropillars positioned at equal distances from a ridge waveguide to enable evanescent coupling of coherent light into the waveguide (see scanning electron microscopy (SEM) images in Fig.~\ref{fig:layout_schematics_and_sem}(c) and (d). Since the coupling efficiency decreases exponentially with the pillar-to-waveguide gap distance according to coupled-mode theory \cite{Roche2024}, devices with suitable gap distances ranging from 100 nm to 900 nm were fabricated and investigated. The structures feature a global Au n-contact and individually addressable p-contacts for each microlaser, allowing selective electrical excitation. Multimode ridge waveguides with two different outcoupling elements —a grating outcoupler and a 2.7 µm micropillar containing QDs— enable efficient light extraction and future resonant excitation schemes without spectral filtering. Additionally, a dedicated p-contact on the waveguide-integrated micropillar allows spectral tuning of the QD emission via the QCSE. 
More details on the device fabrication are provided in the Supplementary Information (SI).

\section{Gap-distance-dependent lasing characteristics of the WGM micropillars}
\label{sec:lasing_characteristics}

We first investigate the emission properties of the monolithically integrated  electrically driven WGM microlasers as function of the gap distance $d_{gap}$. To this end, the fabricated sample was mounted in a He-flow cryostat equipped with a needle probe to study the lasers’ electroluminescence (EL) at 20 K. The sample could be precisely positioned along the X, Y, and Z directions using three high-precision piezo stages. Using a \textit{Keithley 2400 SourceMeter} DC voltage source, this setup enables flexible excitation of single microlasers within the same cooling cycle without reopening the cryostat and before wire-bonding suitable devices for subsequent detailed investigations.

Upon injection of current, each WGM pillar emits laterally, with its light scattering across nearby structures and the sample surface. A NIR 20× micro-objective (NA = 0.4, focal length 200 mm) positioned outside the cryostat collects a portion of the scattered light along the surface-normal direction. The signal is then analyzed using a spectrometer (\textit{Acton SP2750} monochromator 0.75 m focal length, with Si-CCD cameras, spectral resolution of 0.023 nm).

\begin{figure}
    \centering
    \includegraphics[width=0.99\textwidth]{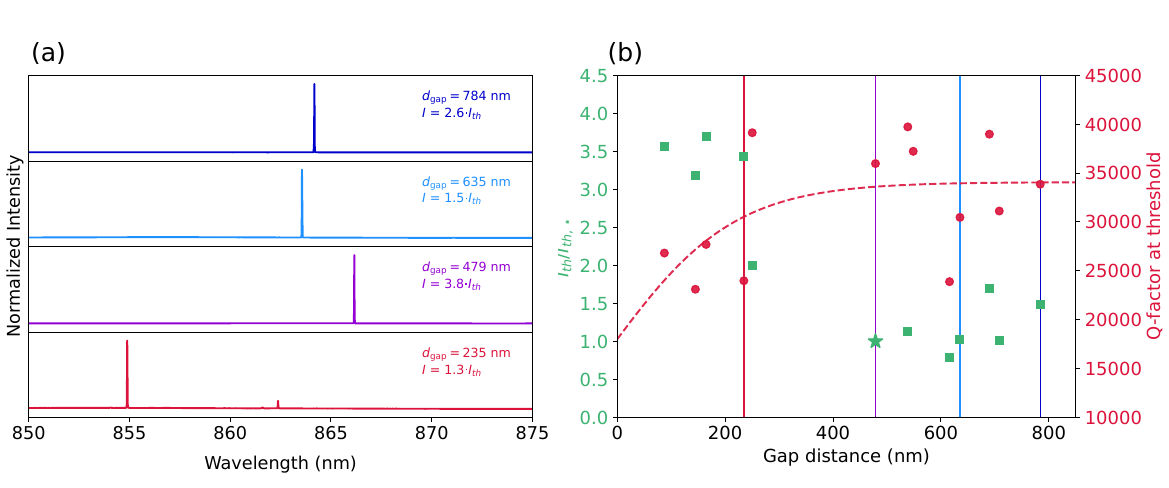}
    \caption{Gap-distance-dependent lasing characteristics of 7.2 µm WGM microlaser pillars. (a) EL spectra of four WGM microlasers measured above their lasing threshold for various gap distances $d_{\mathrm{gap}}$. (b) Relative lasing threshold currents $I/I_{th}$ (green) and Q-factors at threshold (red) of the same 7.2 µm-diameter WGM micropillars as a function of $d_{\mathrm{gap}}$. Vertical lines indicate the structures whose spectra are shown in (a) with the corresponding colors.}
    \label{fig:lasing_characteristics}
\end{figure}

Owing to the p–i–n layer design, the WGM lasers exhibit diode-like current–voltage characteristics, as shown in the SI, Fig.~S2. Under forward bias, the investigated micropillars, all with a diameter of 7.2 µm, display pronounced eEL, characterized by the emission of at least one high-intensity WGM with a narrow linewidth. Fig.~\ref{fig:lasing_characteristics}(a) shows the EL spectra of four WGM micropillar structures with different gap distances $d_{\mathrm{gap}}$ from 235 nm to 785 nm, operated above their respective lasing threshold currents $I_{\mathrm{th}}$. All lasers show pronounced single-mode lasing with emission in the wavelength range between 850 nm and 870 nm as indicated in the figure. 
For the device with $d_{gap} = 235$ nm, an additional mode is present with $\lambda = 862$ nm. The measured side-mode suppression ratio (SMSR) is 17.6 dB at $I = 1$ mA.
Interestingly, this wavelength range is shorter than that of the QD emission band and instead corresponds to the typical emission of the wetting layer \cite{Fricker2022}. We therefore attribute the observation to wetting-layer lasing, rather than QD lasing.

Next, to gain deeper insight into the $d_{\mathrm{gap}}$-dependent emission properties, we extracted the lasing threshold currents and quality factors ($Q$) of 13 WGM microlasers with $d_{\mathrm{gap}}$ ranging from 87 nm to 784 nm by recording injection-current-dependent EL spectra. 
For each EL spectrum, the full width at half maximum (FWHM) was determined by fitting a Voigt profile. The Voigt FWHM is calculated using the approximation $ f_{V} \approx 0.05343 \cdot f_{L} + \sqrt{0.2169 \cdot f_{L}^{2} + f_{G}^{2}} $, where $f_{G}$ and $f_{L}$ denote the Gaussian and Lorentzian FWHMs, respectively, as reported in Ref.\cite{Olivero1977}. The lasing threshold currents were subsequently evaluated by fitting the resulting light–input–output characteristics, as described in detail in the SI.
The resulting lasing threshold currents, normalized to that of the reference device with $d_{\mathrm{gap}}$ = 500 nm (marked by a green star), are plotted as a function of the gap distance in Fig. \ref{fig:lasing_characteristics}(b) (green data points). The reference device is examined in greater detail in the following section. Notably, microlasers with $d_{gap} <$ 200 nm exhibit threshold currents up to 3.5 times higher than those of structures located more than 400 nm away from the ridge waveguide. As the coupling strength between the WGM and the waveguide is expected to increase exponentially with decreasing gap distance, enhanced out-coupling losses are anticipated for structures positioned closer to the waveguide, leading to higher lasing threshold currents for those microlasers\cite{Roche2024}.

The Q-factor is a key parameter for assessing the resonator performance and is experimentally determined as $Q = \lambda / f_V$, where $\lambda$ is the resonance wavelength and $f_V$ is the FWHM of the optical mode. However, since the Q-factor is an intrinsic property of the resonator, its evaluation in microlasers requires particular care, as the measured linewidth depends on the excitation strength. Specifically, below threshold, the linewidth is usually limited by absorption losses, whereas above threshold coherent emission reduces the linewidth. The Q-factor that reflects cavity losses alone -- the so-called \textit{cold-cavity} Q -- corresponds to the transparency condition and is typically calculated by considering the emission linewidth in the vicinity of the lasing threshold \cite{Jones2010}.

Using the previously extracted threshold currents, the corresponding Q-factors at (assumed) transparency were determined for the WGM microlasers from the Voigt-profile fits. The resulting values are plotted in Fig.~\ref{fig:lasing_characteristics}(b) as a function of $d_{\mathrm{gap}}$ (red data points). The measured Q-factors range from approximately ($2\times10^{4}$) to ($4\times10^{4}$), in good agreement with values reported in comparable studies \cite{Jones2010, Limame2024}. The observed spread likely arises from variations in fabrication quality, such as surface roughness, as well as uncertainties in the determination of the lasing threshold. 
Nevertheless, an overall increase in the cavity Q-factor with increasing $d_{\mathrm{gap}}$ is evident, as indicated by red dashed line. This trend is expected, since for small $d_{\mathrm{gap}}$ the \textit{cold-cavity} Q is limited by coupling losses into the waveguide, whereas for larger separations this loss channel is progressively suppressed and the Q-factor approaches that of the uncoupled resonator.
It is expected that the Q-factor evolves with $d_{\mathrm{gap}}$ according to 

\begin{equation}
    Q(d_{\mathrm{gap}}) = \frac{Q_{\mathrm{int}}}{1 + \kappa_{0}e^{-\xi d_{\mathrm{gap}}}}
    \label{eq:Q_gap_function}
\end{equation}

where $Q_{\mathrm{int}}$ is the intrinsic Q-factor of the pillar resonator, $\kappa_{0}$ (dimensionless) the zero-gap pillar-to-waveguide coupling rate and $\xi$ (nm$^{-1}$) the coupling characteristic length \cite{Ding2010, Brooks2021}.
This function was fitted to the measured \textit{cold-cavity} Q-factors from devices with various $d_{\mathrm{gap}}$ and is shown in Fig.~\ref{fig:lasing_characteristics}(b) as a red dashed curve. 
The corresponding fitted parameters are $Q_{\mathrm{int}} = (34~071 \pm 2~444)$, $\kappa_{0} = (0.8 \pm 1.3)$  and $\xi = (0.9 \pm 1.0) \times 10^{-2}$ nm$^{-1}$.
According to the fitted model, the predicted zero-gap Q-factor is approximately $18 \times 10^{3}$.

\section{On-chip excitation of single-photon sources in the waveguide}
\label{sec:on_chip_excitation_of_qds_in_wg}

Next, we focus on the reference device marked by a green star in Fig.~\ref{fig:lasing_characteristics}(b), featuring a nominal $d_{\mathrm{gap}} = 500$ nm. After the pre-characterization, the contact pads of the device were wire-bonded to a chip carrier compatible with a closed-cycle cryostat operated at 4 K. 
The electrical connections enable independent electrical excitation of one WGM micropillar and the simultaneous application of a bias voltage to the aperture contact of the WIM. This allows spectral tuning of the QDs in the active layer via the QCSE.

Luminescence from the WGM lasers and the WIM is collected from the top using a near-infrared-optimized microscope objective (NA = 0.4, 20x magnification) in a confocal configuration. The detection path incorporates a pinhole with a diameter of 30 µm and two lenses with focal lengths of 100 mm and 50 mm, respectively. This configuration yields a spatial selection corresponding to a circular area of approximately 3 µm in diameter on the sample surface (see the corresponding experimental setup schematic in the SI).
By aligning the sample stage with the optical collection path, the aperture system can be used to selectively collect emission either from the outcoupler or from the WIM.

We first investigate the EL of the left microlaser, denoted WGM~L, and its evanescent coupling to the waveguide. Once coupled into the waveguide, the light propagates in both directions and is subsequently redirected normal to the sample plane by the WIM at one end and by the outcoupler at the opposite end.

In the present configuration, the waveguide-coupled emission from WGM~L is analyzed by collecting the signal above the WIM using the previously described spatial filtering with a 3-µm-diameter circular aperture. This filtering ensures that only the waveguide-mediated signal emitted from the WIM is detected, while stray light from WGM~L, for instance scattered at the waveguide surface, is effectively suppressed.

\begin{figure}
    \centering
    \includegraphics[width=1\textwidth]{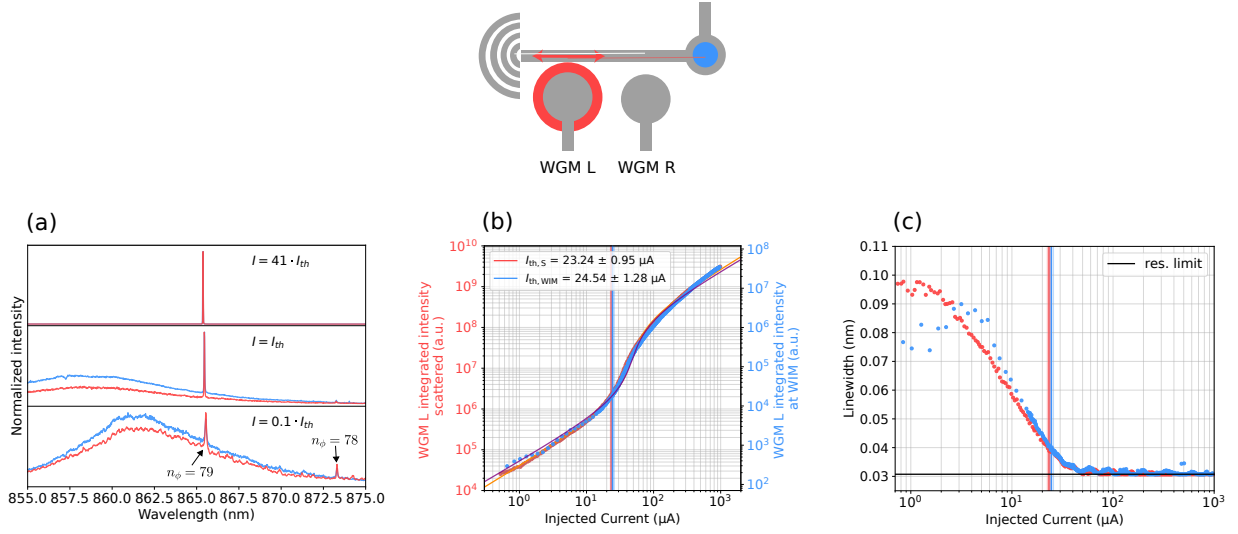}
    \caption{Spatially resolved EL characterization of WGM~L. (a) Overlay of EL spectra for three injection currents, (b) input–output characteristics, and (c) corresponding linewidth evolution of  WGM~L collected at two different detection positions: The red curves correspond to scattered stray light measured directly above WGM~L without spatial filtering. The blue curves correspond to emission from WGM~L evanescently coupled into the waveguide and collected above the WIM using a 3-µm-diameter spatial filter.}
    \label{fig:wgm_l_current_series}
\end{figure}

Figs.~\ref{fig:wgm_l_current_series}(a)–(c) compare the directly scattered emission from WGM~L (red) with the waveguide-mediated signal collected above the WIM (blue).
Fig.~\ref{fig:wgm_l_current_series}(a) presents the EL spectra of WGM~L recorded using the two collection configurations below, at, and above the lasing threshold. The lasing mode is located at approximately 865.5 nm, exhibits a narrow linewidth of 0.039 nm, and receives optical gain from the low-energy tail of the wetting-layer emission. 
An additional side mode is also visible around 873.5 nm. 
Using numerical simulations (see SI), the most prominent mode and the secondary mode are identified as WGMs with azimuthal mode numbers $n_{\phi} = 79$ and $n_{\phi} = 78$ respectively.
The corresponding free spectral range (FSR) is 12.9 meV.

For each injection current, the most prominent mode was fitted with a Voigt profile. The Gaussian width parameter was constrained by a lower bound of 0.013 nm to ensure that the corresponding linewidth remains consistent with the spectrometer resolution of 0.03 nm. Fig.~\ref{fig:wgm_l_current_series}(b) shows the integrated intensity as a function of injection current for both collection methods -- directly above the laser and at the position of the WIM.
Both input–output curves exhibit the characteristic S-shaped behavior of high-$\beta$ microcavity lasers. Importantly, the input–output curve of the waveguide-transmitted signal (blue) shows excellent agreement with that of the micropillar emission, yielding a threshold current of 24 µA, indicated by the vertical blue line.

Both input-output curves are fitted using an adapted laser rate-equation model, as described in the SI. 
The fitted parameters $A$, $B$ and $\beta$ are 
$(42.5 \pm 0.9)$, $(5.4 \pm 0.1)\times10^{-7}$, $(2.00 \pm 0.04)\times10^{-2}$ 
above WGM~L, and
$(43.0 \pm 1.0)$, $(5.3 \pm 0.1)\times10^{-5}$, $(1.96 \pm 0.05)\times10^{-2}$ 
above the WIM, respectively.
The experimental uncertainties of the parameters $A$ and $\beta$ overlap since they represent intrinsic properties of the microlaser. However, the fitted scaling factor $B$ incorporates the photon losses, which differ by several orders of magnitude between the two measurements at different collection points due to the additional losses generated by the waveguide and the outcoupling element.

The two fitted functions yield low lasing thresholds of $(23.24 \pm 0.94)$ µA and $(24.54 \pm 1.27)$ µA (which agree within the experimental uncertainty), measured respectively above the WGM micropillar and above the WIM (see Fig.~\ref{fig:wgm_l_current_series}(b)).
Fig.~\ref{fig:wgm_l_current_series}(c) shows the change in the Voigt profile linewidth $f_v$ as function of the injection current. 
As expected for a structure undergoing the transition to stimulated emission, the emission linewidth narrows significantly from 0.1 nm to 0.03 nm due to the onset of temporal coherence at threshold

In the following, we aim to optically drive the QDs located in the WIM by the evanescently coupled WGM microlaser. The latter one is electrically driven and both the microlaser signal and the QD emission signal are collected above the WIM at the end of the waveguide using the same 3-µm diameter spatial filter as mentioned previously. 

In order to independently quantify the power contributions of the wetting-layer and  the WGM for a specific injected current, both emission peaks are fitted using respective Voigt profiles and are then integrated. The two corresponding integrated intensities are denoted $\Iwl$ and $\Iwgm$. 
The fitted function of spectra corresponding to the injected current above, at and below the lasing current threshold are presented in the SI. 
We denote the total intensity originating from WGM~L in the waveguide as $\Itot = \Iwl + \Iwgm$. 
Furthermore, the ratio of the WGM laser emission power to the total emission power of WGM~L coupled into the waveguide is defined as $\etawgm = \Iwgm/\Itot$. This ratio estimates the contribution of the WGM relative to that of the wetting layer in the excitation of the QDs within the WIM.

\begin{figure}
    \centering
    \includegraphics[width=0.99\textwidth]{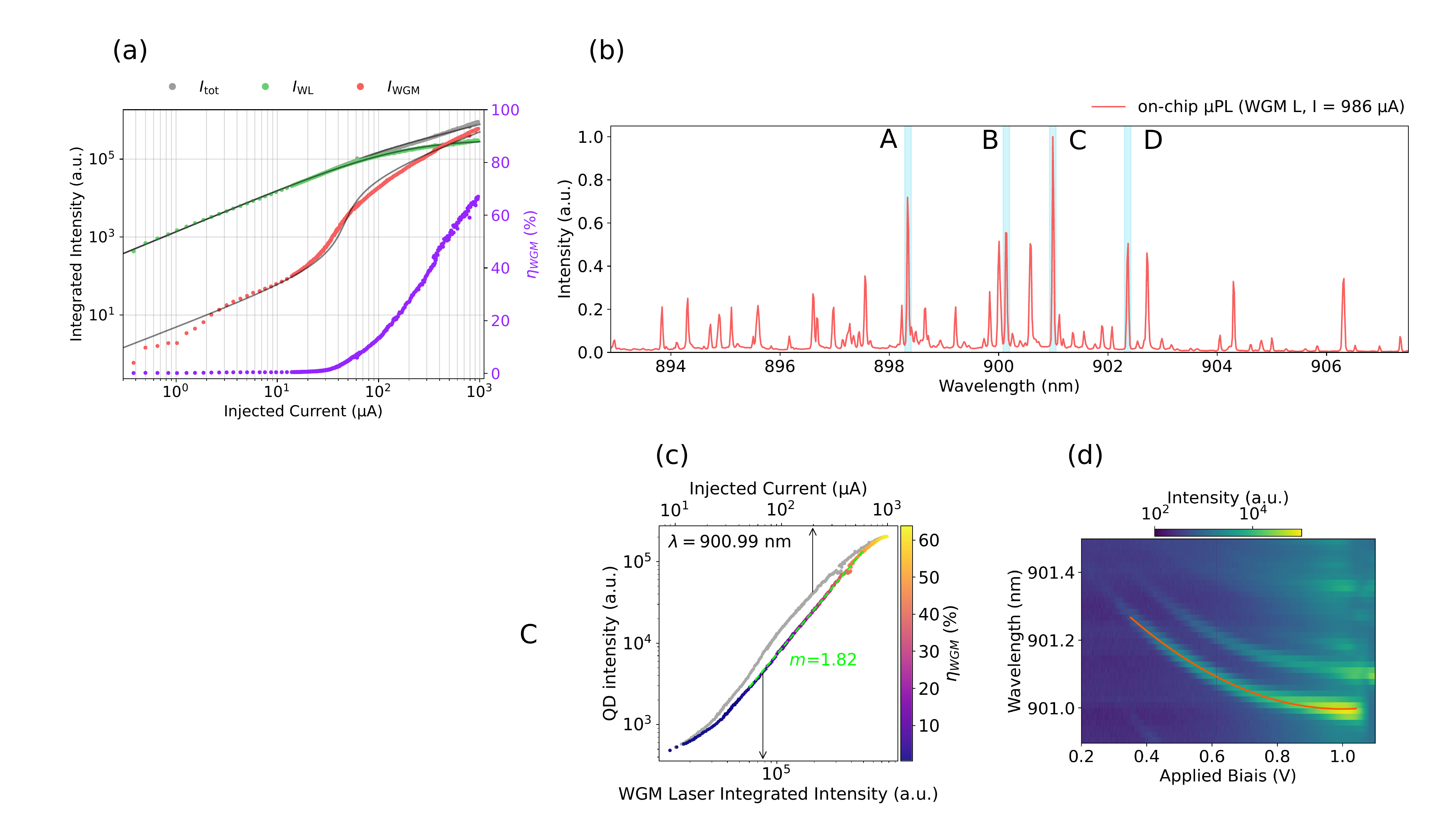}
    \caption{\label{fig:qd_current_series_and_st} Current-dependent emission of the electrically driven WGM~L microlaser and excitation and Stark tuning of QDs in the WIM. (a) Wetting-layer ($\Iwl$), WGM ($\Iwgm$), and total ($\Itot$) integrated intensity of the electrically driven WGM~L microlaser emission as a function of the injection current. The ratio of intensity from the WGM over the total emitted power is given by $\etawgm$. (b) \upl of QDs in the WIM excited by evanescently coupled emission from WGM~L operated at $I = 986$ µA. The measurement was performed with a bias voltage of $V = 1$ V applied to the WIM. Focus on the selected QD emission line C : (a) emission intensity in terms of the injected current in the microlaser and in terms of the integrated intensity of the microlaser emission ($V = 0$ V) and (d) spectral fine-tuning of the emission wavelength via the quantum-confined Stark effect ($I = 700$ µA). 
    All data presented in this figure were collected above the WIM using a 3-µm-diameter spatial filter.}
\end{figure}

Fig.~\ref{fig:qd_current_series_and_st}(a) shows the computed values of $\Iwl$, $\Iwgm$, $\Itot$ and $\etawgm$ for various currents $I$ injected into WGM~L. The WGM intensity $\Iwgm$ in red shows a typical S-shaped curve with a fitted $\beta$-factor of $(1.00 \pm 0.04)\times 10 ^{-2}$.
For injection currents below the lasing threshold, the wetting layer intensity $\Iwl$ in blue is approximately two orders of magnitude higher than the WGM intensity, even when the wetting layer amplitude is lower than the WGM. This is due to the narrow linewidth of the lasing WGM compared to the broad emission spectrum of the wetting layer. Around $I \approx 400$ µA, the WGM power exceeds the wetting layer power and becomes the most important power contribution of the on-chip integrated excitation source, which is associated with $\etawgm >$ 50\%.
The latter increases strongly starting slightly above the lasing threshold and reaches values greater than $60$\% for a maximum injection current of 1 mA.
The computed values of $\Iwgm(I)$ are fitted using the laser rate-equation described in the SI. To fit the wetting-layer intensity $\Iwl(I)$ we developed a phenomenological saturation model $\Iwl(I) = \Iwl^{sat} \cdot \frac{I^{c}}{I^{c} + I_{0}}$ where $\Iwl^{sat}$ is the saturation value of the integrated intensity and $I_{0}$ is the onset of saturation. The fitted functions are shown as black curves in Fig.~\ref{fig:qd_current_series_and_st}(a).

EL from WGM~L coupled into the waveguide is used to excite QDs in the WIM. The corresponding \upl of these QDs is shown in Fig.~\ref{fig:qd_current_series_and_st}(b), where the microlaser WGM~L is operated at an injection current of $I = 986$ µA. 
For comparison, the \upl emission excited by an external continuous-wave (CW) laser is shown in the SI. A bias voltage of 1 V is applied to the WIM which is effective along the whole waveguide structure. 

Four emission lines resulting from the on-chip micropillar excitation are selected and investigated with respect to power dependence and spectral fine-tuning capabilities. 
These emission lines are marked from A to D on Fig.~\ref{fig:qd_current_series_and_st}(b). 
To study the on-chip excitation of the QDs present in the waveguide in more detail, the waveguide was biased at 1 V and the WGM micropillar was driven with various injection currents up to 1 mA. 
Fig.~\ref{fig:qd_current_series_and_st}(c) presents the intensity of the most prominent emission line C as a function of the current injected into WGM~L.
The emission intensity is also shown in terms of the total integrated intensity from the waveguide-coupled microlaser $\Itot(I)$, which is computed as the sum of the two fitted functions $\Iwgm(I)$ and $\Iwl(I)$. 
The color of the plotted dots corresponds to the value of $\etawgm$ (\%), indicating the contribution of the WGM to the overall exciting power. 
This allows us to observe the power dependence of intensity and fit the power parameter $m$ according to $I_{\text{QD}}(I) = c \cdot \Itot(I)^{m}$, where c is a scaling parameter.
The fitted function and the $m$ parameter are both shown in green. 
The power dependence is close to a quadratic dependence which suggests that the observed line corresponds to a double exciton state (XX) \cite{Sk2010,Wijitpatima2024}.
This confirms the successful excitation of the QDs by our on-chip microlaser. 
In order to investigate the spectral tuning capabilities of the QDs in the WIM via the QCSE, the WGM~L excited \upl (at $I = 700$ µA)  was recorded while varying the bias voltage applied to the waveguide from 0 to 1 V (see Fig.~\ref{fig:qd_current_series_and_st}(d) for emission line C). As can be seen in the IV-curve in Fig.~S2 in the SI, this voltage range is below the onset of forward current of the \pin-junction. The resulting QCSE leads to a shift in the QDs emission wavelength of up to 0.35 nm. 

The corresponding positions of the tuned emission lines were fitted using Voigt profiles.
We denote the electric field applied on the QDs $F(V) = (V_{bi} - V) / d$, where $V_{bi} = 1.5 $ V is the measured built-in voltage and $d = 272 $ nm the intrinsic layer thickness.
The energy of a QD $E$ (meV) shifts as function of the applied axial electric field $F$ (kV/cm) according to the expression $E(F) = E_{0} - p F + PF^{2}$, where $E_{0}$ (meV) is the zero-field energy, $p$ (µeV cm/kV) the permanent dipole moment and $P$ (µeV cm$^{2}$/kV$^{2}$) the polarizability \cite{Fry2000}. The three previous parameters were extracted using a quadratic fit on the energy of the emission lines in terms of the applied bias voltage and are summarized in Table~\ref{tab:qd_lines_ST_fit_results}. 
The permanent dipole moment and the polarizability of the different lines are in the same range as reported in previous studies but do not suffice to precisely determine the nature of the excited states alone \cite{Bennett2010, Schnauber2021}.
Nevertheless, this confirms the ability of our device to fine-tune the emission energy of the QD located in the ridge waveguide and in the WIM via the QCSE. This spectral tuning range could potentially be increased by using barrier layers to reduce the tunneling rate \cite{Bennett2010}.
An identical analysis of the other selected emission lines A, B and D are available in Sec.~IX of the SI.

\begin{table}
    \centering
    \begin{tabular}{ccccc}
        \hline
        \textbf{Emission Line} & 
        $\boldsymbol{m}$ & 
        $\boldsymbol{E_{0}}$ \textbf{(meV)} & 
        $\boldsymbol{p}$ \textbf{($\mu$eV cm/kV)} & 
        $\boldsymbol{P}$ \textbf{($\mu$eV cm$^{2}$/kV$^{2}$)} \\
        \hline
        A & $1.33 \pm 0.01$ & $1379.883 \pm 0.004$ & $26.7 \pm 0.3$ & $-0.669 \pm 0.005$ \\
        B & $1.32 \pm 0.01$ & $1377.079 \pm 0.004$ & $30.7 \pm 0.2$ & $-0.768 \pm 0.004$ \\
        C & $1.82 \pm 0.01$ & $1375.832 \pm 0.003$ & $26.8 \pm 0.2$ & $-0.729 \pm 0.004$ \\
        D & $1.75 \pm 0.01$ & $1373.755 \pm 0.003$ & $24.6 \pm 0.2$ & $-0.629 \pm 0.004$ \\
        \hline
    \end{tabular}
    \caption{For each of the selected QD emission lines are determined: the energy at zero electric field $E_{0}$, the permanent dipole moment $p$ and the polarizability $P$.}
    \label{tab:qd_lines_ST_fit_results}
\end{table}

\begin{figure}
	\centering
	\includegraphics[width=0.49\textwidth]{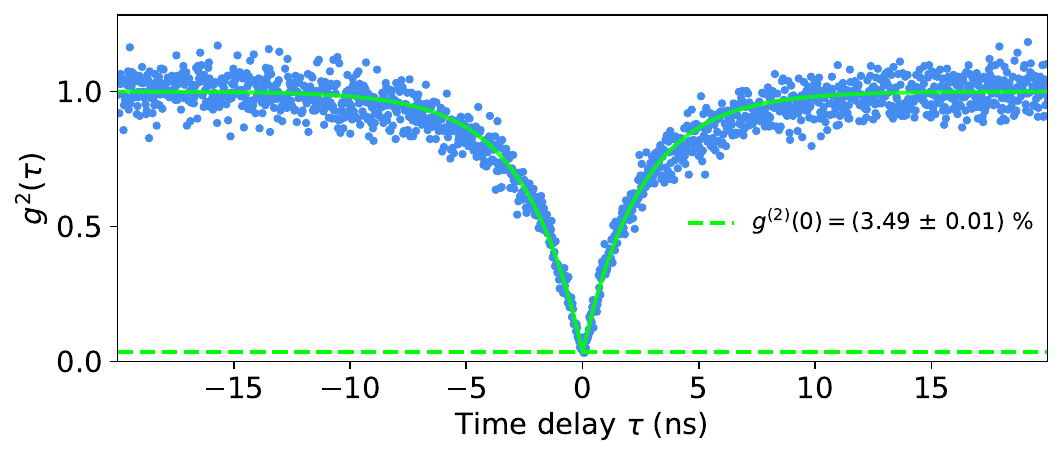}
	\caption{\label{fig:g2}
            Second-order photon correlation measurement of the emission line C ($\lambda = 901.95$ nm) of the WIM under continuous-wave above-band excitation via the WGM~L microlaser which demonstrates clean single-photon emission.}
\end{figure}

Finally, to investigate the quantum nature of emission from the WIM under on-chip WGM excitation, the emission line C located at $\lambda = 901.95$ nm (see Fig.~\ref{fig:qd_current_series_and_st}(c)) was spectrally selected and redirected to a fiber-based Hanbury Brown and Twiss (HBT) setup. A continuous current of 1 mA was injected into the WGM~L microlaser resulting in an above-band, CW on-chip optical excitation of the selected QD. Additionally, a voltage bias of 1 V was applied to the WIM. The resulting normalized second-order photon autocorrelation histogram $g^{(2)}(\tau)$ where $\tau$ is the time delay between two detections is presented in Fig.~\ref{fig:g2}. The correlation histogram shows clear antibunching at zero time delay $\tau = 0$. Fitting the experimental data with $g^{(2)}(\tau) = 1 + b_{0} e^{\frac{-2|\tau|}{\tau_{c}}}$, where $b_0$ is the bunching amplitude and $\tau_{c}$ the coherence time yields $g^{(2)}(0) = 3.49 \pm 0.01$\% \cite{Ulrich2007}. 

These values indicate a good single-photon purity of the measured photon which confirms the quantum nature of the emitter.
We attribute the limits of the purity to uncorrelated emissions such as laser stray light and detection of uncorrelated emission of spectator QDs also located in the WIM. 
The purity could be improved by using narrow spectral filters and by making sure to resonantly excite only a single quantum emitter. 
The measurement of the single-photon purity and the photon indistinguishability under pulsed excitation from the on-chip microlaser is naturally the next step to characterize this type of device.

\section{Conclusion}

In conclusion, we have demonstrated the concept and fabrication of a quantum nanophotonic circuit consisting of a WGM microlaser coupled to a ridge waveguide which excites QDs in a WIM. This architecture represents a compact, electrically driven, on-demand single-photon source suitable for highly functional IQPCs. The electrically contacted WGM micropillar exhibits clear lasing behavior with the QD wetting layer as the gain medium. By systematically analyzing the pillar–waveguide coupling, we determined the lasing threshold current of 23.24 µA and found that devices with a pillar-to-waveguide gap below 300 nm exhibit more than a two-fold increase in threshold current compared to devices with larger gaps, highlighting the impact of evanescent coupling on the microlaser dynamics. A device with WGM micropillars positioned at a distance of 500 nm from the waveguide was investigated in detail, revealing a lasing threshold below 100 µA and a progressive linewidth narrowing down to the instrumental resolution limit.

Exploiting evanescent coupling from the WGM microlaser, we achieved on-chip \upl excitation of QDs embedded in the WIM. Through a detailed analysis of the current-dependent emission spectra coupled into the waveguide, we investigated multiple QD transitions via their on-chip power dependence. Furthermore, we demonstrated spectral tuning of these emission lines by up to 0.35 nm via voltage-controlled Stark shifts applied to the waveguide structure. For the brightest QD transition, we measured a photon antibunching value of $g^{(2)}(0) = (3.49 \pm 0.01)\%$ under CW electrical excitation via the integrated WGM microlaser, confirming high-purity single-photon emission under on-chip excitation.

\section*{Acknowledgments}
The research leading to these results received funding from the German Research Foundation (DFG) via Grant RE2974/29-1 and from Berlin Quantum via the SQALE project.

\section*{Supporting information}

Supporting Information Available: 
Extended device description, experimental setup, IV-curve of the \textit{p-i-n}-doped structures, laser rate-equation model, evanescent coupling between WGM R and waveguide, numerical modelling of WGM emission, wetting layer and WGM Voigt profile fits, comparison between on-chip and off-chip QD excitation in the WIM and additional QD emission lines analysis. This material is available free of charge via the Internet at http://pubs.acs.org

\newpage

%\bibliography{export}

\providecommand{\latin}[1]{#1}
\makeatletter
\providecommand{\doi}
  {\begingroup\let\do\@makeother\dospecials
  \catcode`\{=1 \catcode`\}=2 \doi@aux}
\providecommand{\doi@aux}[1]{\endgroup\texttt{#1}}
\makeatother
\providecommand*\mcitethebibliography{\thebibliography}
\csname @ifundefined\endcsname{endmcitethebibliography}
  {\let\endmcitethebibliography\endthebibliography}{}

\end{document}

\maketitle

This Supporting Information provides additional experimental and theoretical details complementing the main manuscript. It includes a detailed description of the studied device, current–voltage characterization of the two electrically isolated \textit{p-i-n}-doped structures (the electrically contacted whispering-gallery-mode (WGM) micropillar and the waveguide-integrated micropillar (WIM)), and the laser rate-equation model used to analyze input–output characteristics and determine lasing thresholds. 
Further, we present a study of evanescent coupling between the WGM resonator and the waveguide, along with details of the numerical simulations of the WGMs. We also provide representative Voigt-profile fits to wetting-layer and WGM emission spectra used to extract integrated emission intensities as a function of injection current. 
Finally, we include a comparison of on-chip and off-chip optical excitation of quantum dots (QDs) in the WIM, as well as additional data on current-dependent emission and Stark tuning of excited QD emission lines.

\section{Extended device description}
\label{supp::extended_device_description}

Each fabricated device is composed of two electrically contacted WGM micropillars with a diameter of 7.2 µm. The two pillars are located at the same pillar-to-waveguide gap distance $d_{gap}$ from the ridge waveguide and are identical in design. Their purpose is to evanescently couple the emitted coherent light to the neighboring waveguide. According to coupled-mode theory (CMT) and the previous work studying the complete three-dimensional coupled system using finite element method (FEM) simulations \cite{Roche2024}, the coupling efficiency of the WGM resonator decreases exponentially with the gap distance. For this reason, several devices only differing by $d_{gap}$ ranging from 100 nm to 900 nm were fabricated. 

The backside of the sample features a global Au n-contact, while each WGM microlaser is equipped with an individual p-contact connected to a larger (200 µm $\times$ 200 µm) contact pad. This configuration enables selective electrical excitation of individual WGM micropillars using a needle-probe contact system inside the used He-flow cryostat, providing a flexible and convenient addressing scheme. The ridge waveguides have a width of 920 nm, supporting multimode propagation. To collect the guided optical signal without cleaving the sample, two different outcoupling elements are implemented at the waveguide terminations. At one end, a grating outcoupler with a pitch of (1185 $\pm$ 37) nm and a ridge width of approximately (833 $\pm$ 37) nm is used to extract the guided light vertically. At the opposite end, a 2.7 µm diameter micropillar is integrated into the waveguide. This structure is designed to enhance the spontaneous emission rate of photons emitted by an embedded QD via the Purcell effect. In addition, perspectively it enables excitation and collection of the QD emission through two distinct, non-overlapping optical modes, thereby eliminating the need for spectral filtering during resonant excitation \cite{Huber2020}.

In addition to the p-contacts of the WGM microlasers, a dedicated p-contact with a 1.7-µm diameter aperture is fabricated on top of the WIM. This contact enables the application of a vertical electric field at the position of the QDs in the active layer, allowing their emission to be spectrally fine-tuned via the QCSE. The intrinsic layer containing the QDs of the microlasers is electrically isolated from the intrinsic layer of the waveguide by a separation distance $d_{\mathrm{gap}}$, which ensures electrical isolation while simultaneously allowing efficient evanescent coupling of the WGM into the waveguide.

\section{Experimental Setup Schematic}
\label{supp:experimental_setup_schematic}

\begin{figure}[H]
    \centering
    \includegraphics[width=0.99\textwidth]{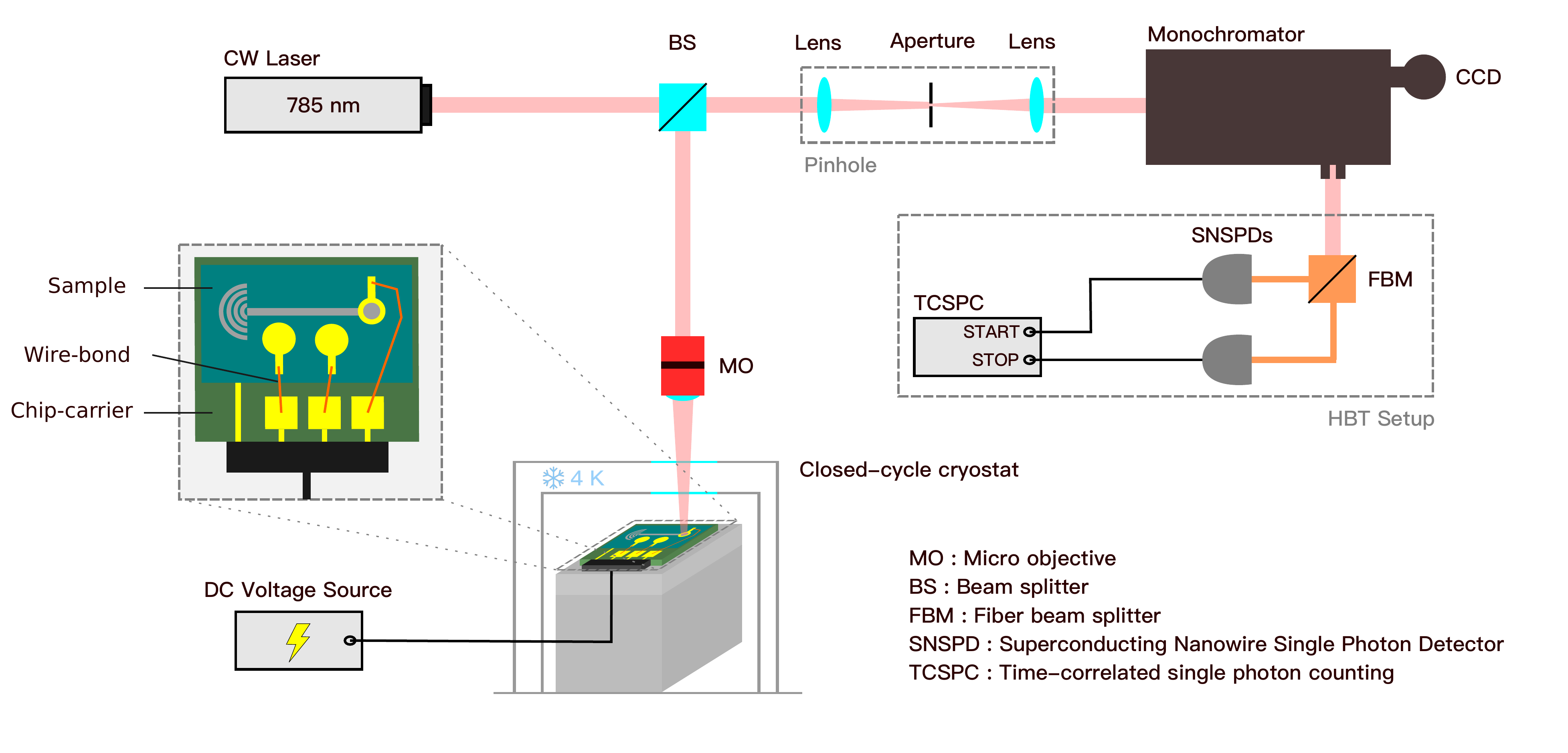}
    \caption{Schematic of the experimental setup used in Sec.~IV.}
    \label{fig:setup_schematics}
\end{figure}

Fig.~\ref{fig:setup_schematics} presents a schematic of the experimental setup used for the optical characterization of our devices.
The contact pads of the two WGM micropillars and the WIM are wire-bonded to gold pads on a chip-carrier, to which the sample is attached using electrically and thermally conductive silver paste. Each electrical channel of the chip-carrier is connected to cryogenic-temperature-tolerant cables, which are routed through the feed-through of the cryostat and connected to a DC voltage source located externally.

The chip-carrier is mounted on a gold finger using screws and silver paste. This assembly is movable along three axes via precise piezo-motor-driven stages. The closed-cycle cryostat is operated at 4 K.

The collection signal is gathered using a micro-objective (NA = 0.4, 20× magnification) positioned above the cryostat. The collected signal is then directed to a 90:10 beam splitter and subsequently to a pinhole system for spatial filtering. This system consists of two lenses with focal lengths of 100 mm and 50 mm, with an aperture placed between them. The optical setup enables spatial selection of a 3-µm-diameter area on the sample surface, allowing measurement of the signal emitted solely from the WIM or the outcoupler.

The filtered signal is spectrally resolved using a monochromator. By default, the signal is directed to and measured by a charge-coupled device (CCD) camera attached to the monochromator to acquire the spectrum. To measure the photon correlation of a QD emission line, the mirror inside the monochromator is flipped, and the side slit is used as a notch filter. The spectrally selected line is then coupled into a fiber beam splitter. The two outputs of the beam splitter are directed to independent superconducting nanowire single-photon detectors (SNSPDs). A time-correlated single-photon counting (TCSPC) module records start and stop events each time the respective SNSPD detects a photon.
This Hanbury Brown and Twiss (HBT) setup allows the measurement of the photon correlation function of a specifically selected QD emission line.

Additionally, an off-chip continuous-wave (CW) laser emitting at 785 nm is used, as described in the previous section. This laser shares the beam splitter and micro-objective of the collection path for excitation.

\section{Current-voltage characteristics of the pin-doped structures}
\label{supp:iv_curves}

\begin{figure}[htbp]
    \centering
    \includegraphics[width=0.49\textwidth]{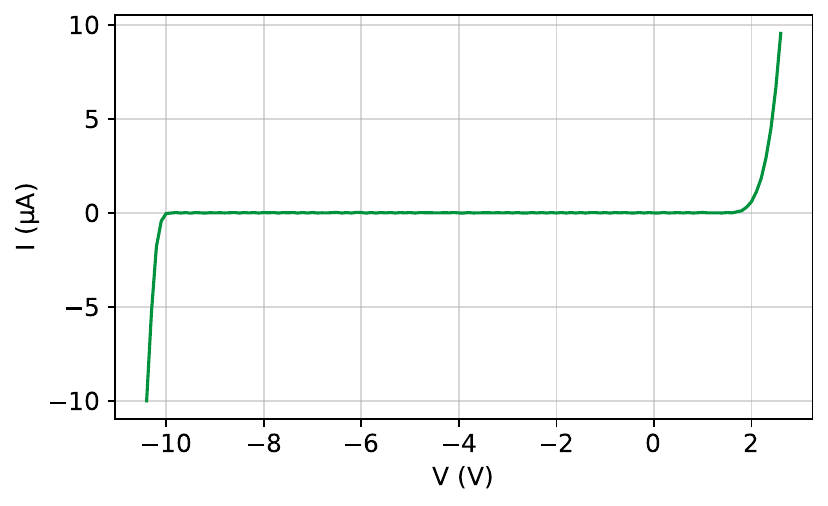}
    \caption{IV curve measured for the 7.2 µm diameter WGM~L pillar with 500 nm gap distance from the waveguide.}
    \label{fig:IV_WGM_pillar_contact}
    \includegraphics[width=0.49\textwidth]{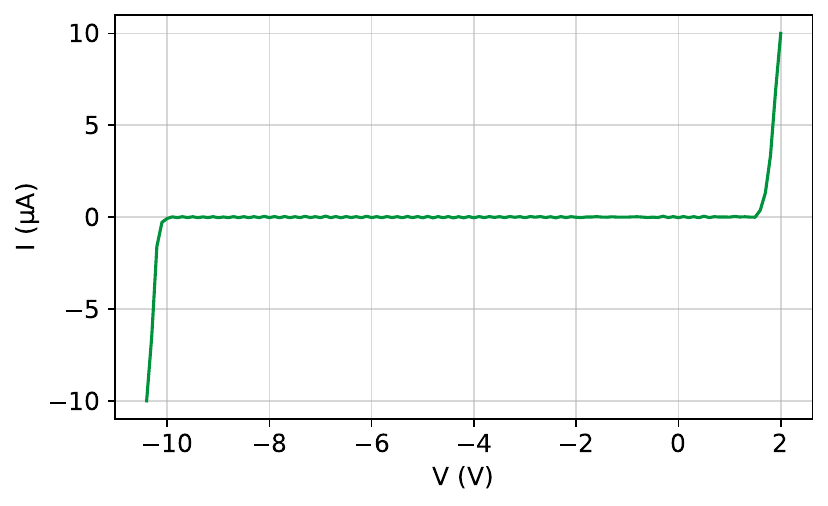}
    \caption{IV curve measured for the WIM.}
    \label{fig:IV_ST_contact}
\end{figure}

Fig.~\ref{fig:IV_WGM_pillar_contact} shows the IV curve measured for the WGM~L pillar with a diameter of 7.2 µm and pillar-to-waveguide gap distance of $d_{gap} = 500$ nm. Fig.~\ref{fig:IV_ST_contact} shows the IV curve measured for the WIM used for spectral tuning the QD via the quantum confined Stark effect (QCSE). Both IV curves reveal the characteristic exponential turn-on of a \textit{p–i-n} junction diode, with a threshold voltage of approximately 1.5 V indicating the onset of efficient carrier injection.

\section{Laser rate-equation model}
\label{supp:laser_rate_equation_model}

The input-output curves of the studied microlasers were fitted using the adapted laser rate-equation model \cite{Bjork1991, Andreoli2021, Shih2023}
\begin{equation}
    I(A, B, \beta) = \frac{A q \gamma}{\beta} \left[  \frac{B p}{1 + B p} (1 + \hat{\xi}\beta)(1 + \beta B p ) - \hat{\xi}\beta^{2} B p \right], 
    \label{eq:laser_rate_equation}
\end{equation}
where $I$ is the injection current (A), 
$p$ is the number of output photons (dimensionless), 
$B$ is the output scaling factor (dimensionless),
$q$ is the electron charge (C), 
$A$ is the current injection scaling factor (dimensionless), 
$\gamma = \frac{2\pi c}{\lambda Q}$ is the cavity decay rate ($s^{-1}$),
$\beta$ is the spontaneous emission coupling factor (dimensionless). 
The parameter $\hat{\xi}$ is defined by $\hat{\xi} = \frac{n_{0}}{\tau_{sp}\gamma}$ where 
$n_{0}$ is the number of excitons at transparency and $\tau_{sp}$ is the spontaneous emission lifetime ($s$).
The parameters $A$, $B$, and $\beta$ are fitted to the input-output data $(I, p)$ while the other parameters are held constant.
The lasing threshold current is considered here as the injection current at which the mean photon number in the lasing mode is unity.
% the active medium becomes transparent and is expressed from the previously fitted parameters using 

\begin{equation}
    I_{\mathrm{th}} = \frac{Aq\gamma}{2 \beta}(1 + \beta + \hat{\xi}\beta  (1 - \beta)).
    \label{eq:threshold_equation}
\end{equation}

The corresponding uncertainty is computed using the expression

\begin{equation}
    \Delta I_{\mathrm{th}}(A, B, \beta)  = 
                \Delta A \cdot | \frac{I_{th}}{A} |  + 
                \Delta \beta \cdot 
                |
                     \frac{Aq\gamma}{2 \beta^{2}}  (1 + \xi)
                |
\end{equation}
where $\Delta A$ and $\Delta \beta$ are the respective uncertainties of the $A$ and $\beta$ parameters.

\section{Evanescent coupling between WGM~R and waveguide}
\label{supp:wgm_r_study}

\begin{figure}[H]
    \centering
    \includegraphics[width=1\textwidth]{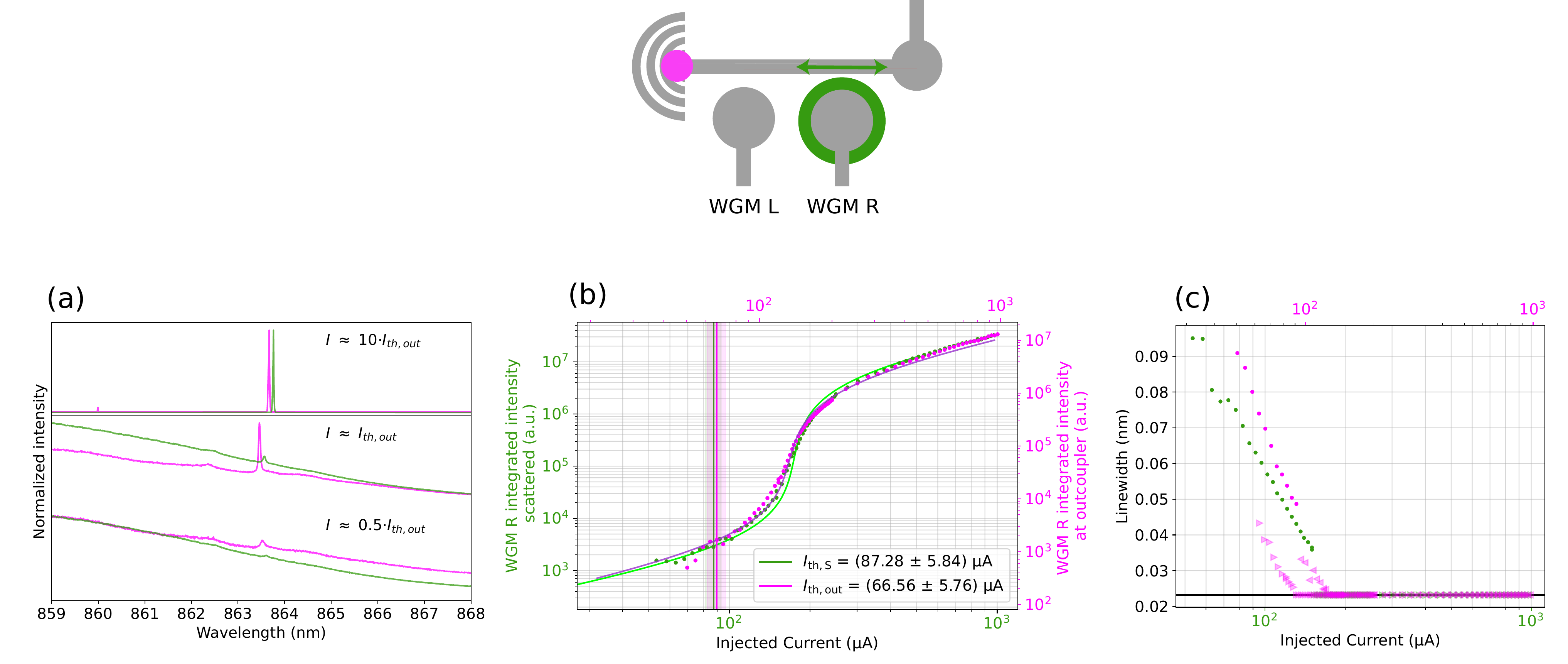}
    \caption{(a) Overlay of EL spectra, (b) input–output curves, and (c) linewidth reduction of WGM~R, collected at two different positions. The green curves correspond to scattered stray light measured directly above WGM~R without spatial filtering. The pink curves correspond to emission from WGM~R evanescently coupled into the waveguide and collected above the outcoupler using a 3-µm-diameter circular spatial filter.}
    \label{fig:wgm_r_current_series}
\end{figure}

Next, we study the optical properties of WGM~R and its coupling into the waveguide. Again, as shown in Fig.~\ref{fig:wgm_r_current_series}(a), (b) and (c), the emission properties were investigated for different injection currents and the emissions were collected at two different collection spots at 20 K. Firstly, the scattered stray light emission from the micropillar was collected directly above WGM~R without spatial filtering via the pinhole and corresponds to the green curves.
The emission coupled into the waveguide was collected above the outcoupler by using the same 3-µm diameter circular spatial filtering and is represented by the pink curves. In this measurement configuration, WGM~L blocks the lateral emission from WGM~R, which could otherwise reach the outcoupler directly through the air.

Fig.~\ref{fig:wgm_r_current_series}(a) shows the EL spectra of WGM~R for the two different collection methods, below, at, and above lasing threshold of  ($66.56 \pm 6.76$) µA. 
% The threshold current is notably higher than for WGM~L which can be explained by XX. 
The lasing mode is spectrally close to that of the neighboring pillar, as expected, since both pillars have the same nominal pillar diameter. The 1.7 nm spectral difference between the laser modes of WGM~L and WGM~R can be attributed to fabrication imperfections that lead to a slight difference in the effective cavity diameter. 
By assuming that WGM~L has a diameter of 7.2 µm and by using that $\mathrm{d}E_{\text{WGM}}/\mathrm{d}d = - E / d$, we estimate the diameter difference between the two cavities to be approximately 15 nm.

The integrated intensity in terms of the injected current is presented in Fig.~\ref{fig:wgm_r_current_series}(b) for the two different collection methods. The input-output curves present the characteristic S-shape, which is typical for microlasers. 
The rate equation in Eq.~\ref{eq:laser_rate_equation} was then fitted to the input-output curve. 
The fitted parameters $A$, $B$, $\beta$ are $(4.7 \pm 0.2)$, $(60 \pm 2)\times10^{-5}$, $(10.0 \pm 0.4)\times10^{-4}$ from the outcoupler and $A$, $B$, $\beta$ are $(2.9 \pm 0.1)$, $(34 \pm 1)\times10^{-5}$, $(4.4 \pm 0.1)\times10^{-4}$ above WGM~R.
The lasing threshold current was then extracted from the fitted parameters using Eq.~\ref{eq:threshold_equation}.

The two measurements taken at different locations were fitted again using a single Voigt profile for low current values. Slightly above the laser threshold, the lasing mode splits into two spectrally close but distinct modes. This phenomenon has been observed before and can be explained by the presence of a subwavelength scatterer inside or outside the resonator cavity, possibly introduced by etching. Such a scatterer can break the rotational symmetry of the resonator and thus lifts the degeneracy of the clockwise and counterclockwise WGMs, resulting in a slight energy difference between the two modes \cite{Borselli2004, Jones2010}. The overlapping split modes were fitted using a suitable double Voigt profile.

The linewidth of the single Voigt profile is shown as a circle in Fig.~\ref{fig:wgm_r_current_series}(c). The linewidth of the left Voigt component is indicated by a left-pointing triangular marker, while the linewidth of the right Voigt component is indicated by a right-pointing triangular marker. For both collections we observe a strong linewidth reduction, until reaching the spectrometer resolution limit. This confirms the lasing behaviour of the structures. Moreover, a significant splitting of the lasing mode is observed around the lasing threshold. 

We attribute the difference in lasing threshold currents $(23.24 \pm 0.94)$ µA and (87.28 $\pm$ 5.84) µA between the two neighboring pillars WGM~L and WGM~R to the difference in Q-factor of their cavity which are 22~300 and 13~100 for WGM L and WGM~R respectively. 

\section{Numerical simulations}
\label{supp:numerical_simulations}

To identify the modes observed in the emission spectrum of WGM L presented in the main text, we performed numerical FEM simulations using the software \textit{JCMsuite}. Exploiting the cylindrical symmetry of the micropillar geometry, the system was reduced to a two-dimensional cross-section, thereby significantly reducing the computational overhead. The time-harmonic, source-free Maxwell's equations were solved in cylindrical coordinates with appropriate Bloch-vector boundary conditions, yielding eigenpairs
($\lambda$, $\mathbf{E}$), where
$\lambda$ denotes the resonant wavelength and $\mathbf{E}$ the associated electric field distribution.

Simulating WGMs in the spectral vicinity of the experimentally observed modes across a range of pillar diameters, we identify the following assignment as the most consistent with the measurements: for a pillar diameter of 7180 nm, the predicted resonant wavelength of the WGM with azimuthal mode number $n_{\phi}$ = 79 is in excellent agreement with the dominant experimental mode, while the mode with $n_{\phi}$ = 78 lies within 1.5 nm of the secondary mode. 
We attribute the residual discrepancies between simulation and experiment to three potential factors: uncertainty in the experimentally determined pillar diameter, limited accuracy of the low-temperature refractive indices used in the simulation, and perturbative effects of the waveguide on the micropillar emission spectrum.

\section{Wetting layer and WGM Voigt profile fits}
\label{supp:wl_wgm_fit_examples}

\begin{figure}[H]
    \centering
    \includegraphics[width=0.99\textwidth]{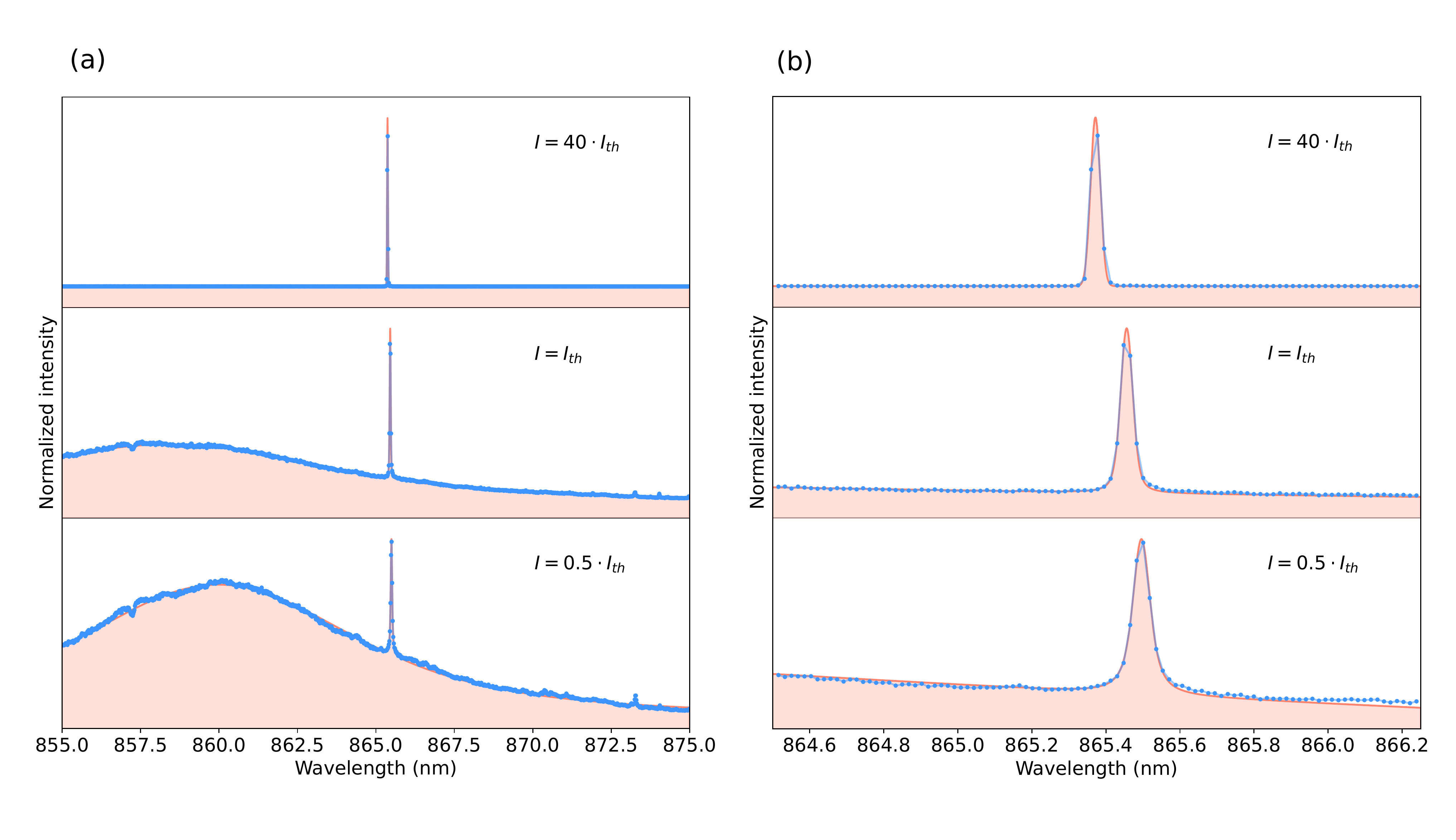}
    \caption{(a) Wide and (b) close-up emission spectra of WGM~L electrically driven above, at, and below lasing threshold and collected above the integrated pillar with a 3-µm diameter spatial filter. The corresponding fit is plotted in red and matches the broad WL emission and the sharp WGM.}
    \label{fig:WGM_WL_fit_reconstruction}
\end{figure}

In order to estimate the total power emitted by the integrated microlaser WGM~L during the on-chip excitation of the QDs in the waveguide in Sec.~IV, the spectra of the microlaser driven by various injected currents were fitted. The wetting-layer (WL) broad emission was fitted using a first Voigt profile and the lasing WGM with a second Voigt profile. 
Fig.~\ref{fig:WGM_WL_fit_reconstruction} shows the emission spectra (blue curve), the reconstructed fitted function (red curve) as well as the area under the fitted spectrum corresponding to the integrated intensity (red). 

\section{Comparison of on-chip and off-chip QD excitation in the WIM}
\label{supp:comparison_on_off_chip_µPL}

\begin{figure}[H]
    \centering
    \includegraphics[width=0.8\textwidth]{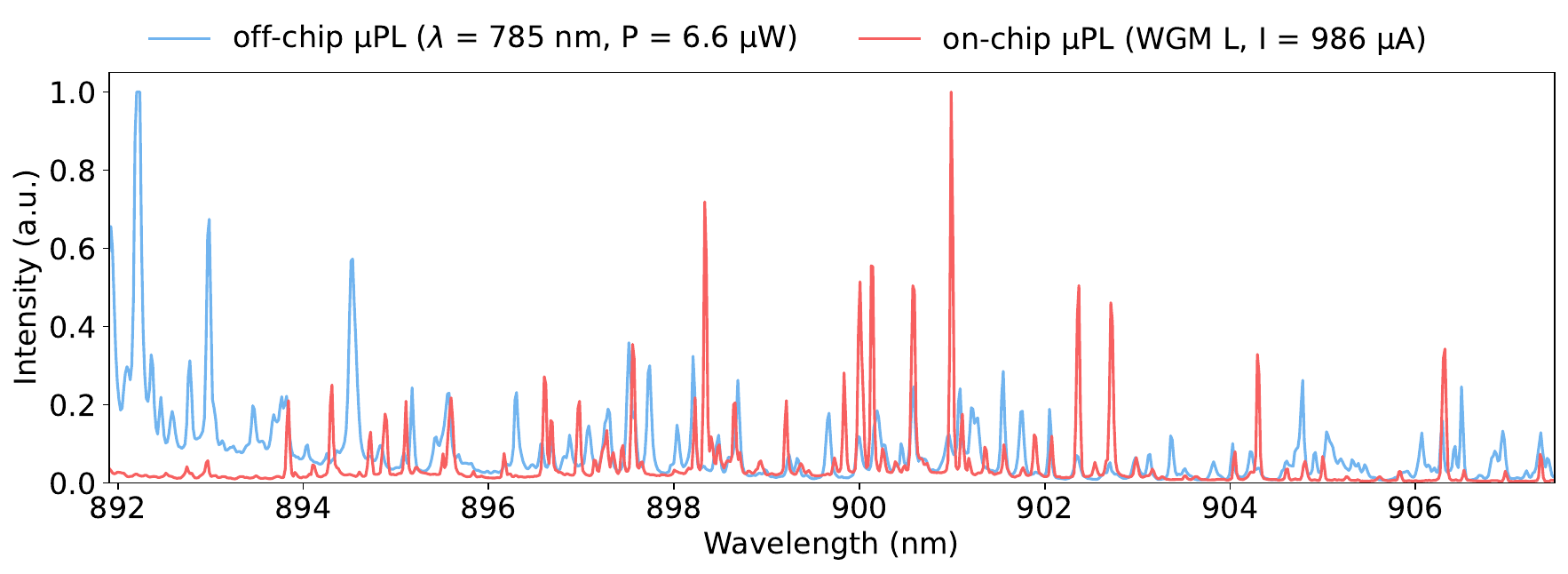}
    \caption{Comparison of the emission spectrum from the QDs located in the waveguide from on-chip and off-chip excitation.}
    \label{fig:QDs_uPL_comparison}
\end{figure}

Fig.~\ref{fig:QDs_uPL_comparison} presents the comparison between the emission signal from QDs in the WIM excited by the electrically driven on-chip WGM microlaser WGM~L ($I = 986$ µA) in red and the off-chip laser focused on the integrated pillar ($\lambda = 785$ nm and $P = 6.6$ µW).
Not all the emission lines are present on both spectra, which can be explained by the fact that the on-chip microlaser excites all the QDs present between WGM~L and the integrated pillar, whereas the off-chip laser was focused on the integrated pillar, thus only exciting QDs present inside the integrated pillar. 

\section{Current-dependent emission and Stark-Tuning of additional on-chip excited QD emission lines}
\label{supp:additional_qd_lines}

\begin{figure}[H]
    \centering
    \includegraphics[width=0.8\textwidth]{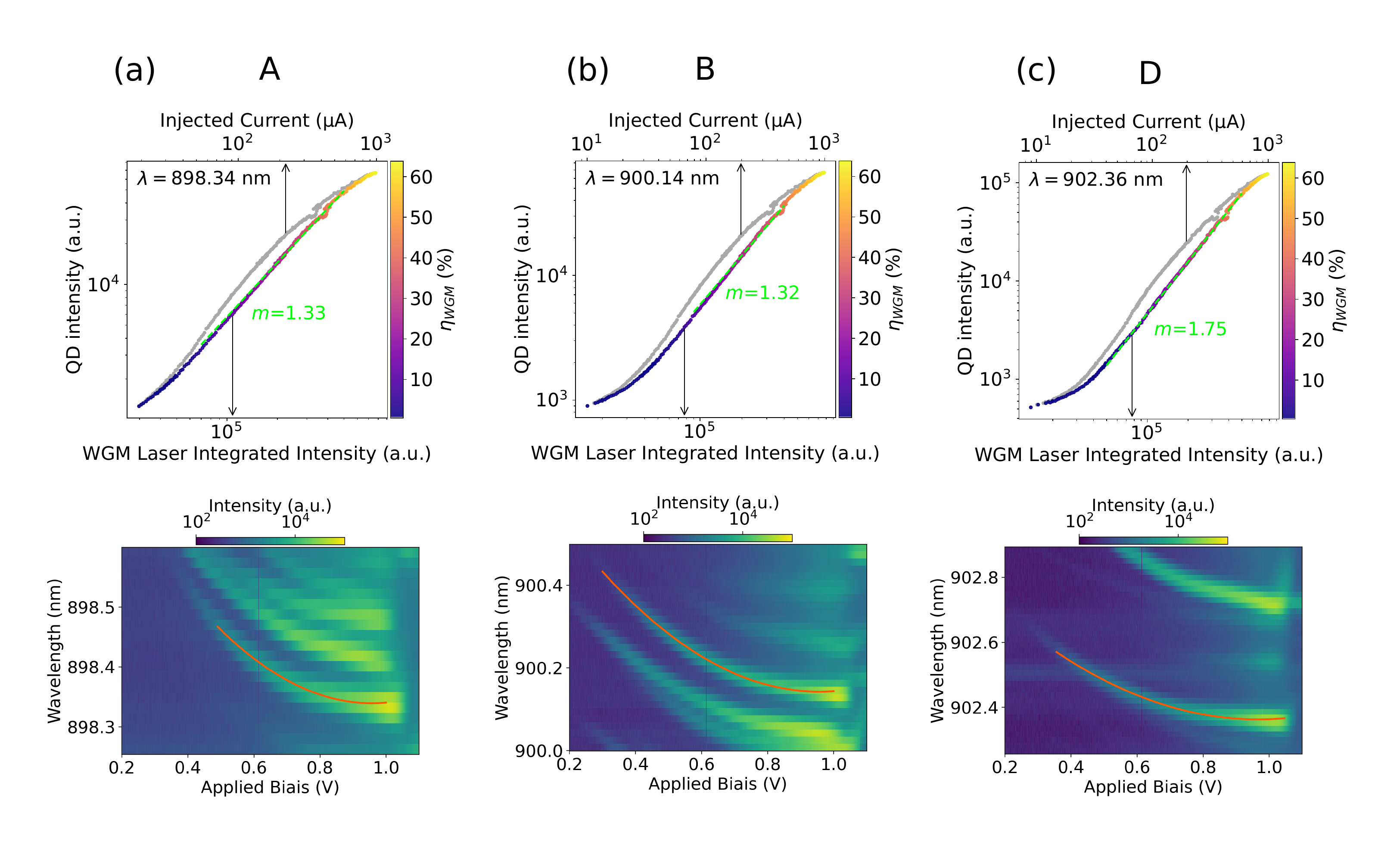}
    \caption{Current-dependent emission and Stark-Tuning of QD emission lines A, B and D.}
    \label{fig:additional_qd_lines}
\end{figure}

We present here the analysis of the three additional QD emission lines. The excitation and the collection was performed in the same conditions as for the emission line C presented in the main text. 
Figs.~\ref{fig:additional_qd_lines}(a), (b) and (c) show the current-dependent emission (top) and the Stark-Tuning of the emission lines A, B and D respectively. 
The fitted function and the $m$ parameter are both shown in green on each plot. Depending on whether the power dependence is close to a linear dependence or to a quadratic dependence, the emission lines can be categorized into two distinct categories, which can be attributed to the characteristics of an exciton state (X) or a biexciton state (XX), respectively \cite{Sk2010,Wijitpatima2024}. This suggests that the emission lines A and B originate from an excitonic state, whereas D from a biexcitonic state.

\clearpage

%\bibliography{export}

\providecommand{\latin}[1]{#1}
\makeatletter
\providecommand{\doi}
  {\begingroup\let\do\@makeother\dospecials
  \catcode`\{=1 \catcode`\}=2 \doi@aux}
\providecommand{\doi@aux}[1]{\endgroup\texttt{#1}}
\makeatother
\providecommand*\mcitethebibliography{\thebibliography}
\csname @ifundefined\endcsname{endmcitethebibliography}
  {\let\endmcitethebibliography\endthebibliography}{}